\documentclass{kapproc} 

\RequirePackage{graphicx}%
\RequirePackage{epsf}%
\input psfig.sty

\upperandlowercase
\let\footnote\savefootnote
\let\footnotetext\savefootnotetext 
 
\def\solmas{{M$_\odot$}}
\def\simless{\mathbin{\lower 3pt\hbox
   {$\rlap{\raise 5pt\hbox{$\char'074$}}\mathchar"7218$}}}   
\def\simgreat{\mathbin{\lower 3pt\hbox
   {$\rlap{\raise 5pt\hbox{$\char'076$}}\mathchar"7218$}}}   
\def\etal{{\rm et al.}}

\def\solmas{{M$_\odot$}}
\def\solm{{M_\odot}}

\def\ms {M_*}

\def\Rtidal {R_{\rm tidal}}

\def\menc {M_{\rm enc}}
\def\Rclus {R_*}
\def\Rbh {R_{\rm BH}}

\kluwerbib 

\begin{document}


\articletitle{Competitive Accretion and the IMF}


\chaptitlerunninghead{Competitive accretion and the IMF}



 \author{Ian A. Bonnell}
 \affil{School of Physics and Astronomy, University of St Andrews, North Haugh, St Andrews, KY16 9SS, UK}
 \email{iab1@st-and.ac.uk}





 \begin{abstract}
 
 Competitive accretion occurs when stars in a cluster accrete from a
 shared reservoir of gas. The competition arises due to the relative attraction
 of stars as a function of their mass and location in the cluster.
 The low relative motions of the stars and gas in young, gas dominated clusters
 results in a tidal limit to the accretion whereas in the stellar dominated cluster cores, the high relative velocities  results in Bondi-Hoyle accretion.
 The combination of these two accretion processes produces
a two power-law IMF with  $\gamma \approx -1.5$, for low-mass 
stars which accrue their mass in the gas dominated regime, and  
 a steeper, $\gamma\approx -2.5$, IMF for higher-mass stars that form in the core of a cluster. Simulations of the fragmentation 
and formation of a stellar cluster show that the final stellar masses, and IMF, are due to 
competitive accretion. Competitive accretion 
also naturally results in a mass segregated cluster and in a direct 
correlation between the richness of a cluster and the mass of the most massive star therein.
The {\sl knee} where the IMF slope changes occurs  near  the Jeans mass of the system.
 \end{abstract}

\section{Introduction}
Star formation is a dynamical process where most stars form in a clustered
ennvironment (Lada \& Lada~2003; Clarke, Bonnell \& Hillenbrand~2000). In
such an environment, stars and gas move in their combined potential on timescales
comparable to the formation time of individual stars. Furthermore, the initial fragmentation
of a molecular cloud is very inefficient (eg., Motte \etal~1998), such that the youngest clusters are dominated by their
gas content. In such an environment, gas accretion can contribute significantly to the final mass of a star.

Stellar clusters are found to be mass segregated even from the youngest ages. The location
of massive stars in the cores of clusters cannot be explained by dynamical mass segregation as the
systems are too young (Bonnell \& Davies~1998). A Jeans mass argument  also fails as the Jeans mass in the core should be lower then elsewhere in the cluster. Mass segregation
is a natural outcome of competitive accretion due to the gas inflow to the centre of the cluster potential.

\section{Accretion in stellar clusters: two regimes}

In a series of numerical experiments , we investigated the dynamics of accretion in gas-dominated stellar
clusters
(Bonnell \etal~1997,~2001a). In the initial studies of accretion in small stellar clusters, we found
that the gas accretion was highly non-uniform with a few stars accreting significantly more than
the rest. This occurred as the gas flowed down to the core of the cluster and was there
accreted by the increasingly most-massive star. Other, less massive stars were ejected
from the cluster and had their accretion halted (see also Bate, Bonnell \& Bromm 2002).
In a follow-up study investigating accretion in clusters of 100 stars, we discovered two
different physical regimes (Bonnell \etal~2001a) resulting in different accretion radii,
$R_{\rm acc}$,  for the mass accretion rate
\begin{equation}
M_{\rm acc} = \rho v \pi R_{\rm acc}^2,
\end{equation}
where $\rho$ is the local gas density and $v$ is the relative velocity of the gas.
Firstly, in 
the gas dominated phase of the cluster, tides limit the accretion as the relative
velocity between the stars and gas is low. The tidal radius, due to the star's position in cluster potential, 
\begin{equation}
\Rtidal \approx 0.5  \left({\ms \over \menc}\right)^{1\over 3} \Rclus,
\end{equation}
is then smaller than the traditional Bondi-Hoyle radius and
determines the accretion rate. Accretion in this regime naturally results in a mass
segregated cluster as the accretion rates are highest in the cluster core.

Once accretion has increased the mass of the stars, and consequently decreased the
gas mass present, the stars begin to dominate the stellar potential and thus virialise.
This occurs first in the core of the cluster where higher-mass stars form due
to the higher accretion rates there.The relative velocity between the gas and stars is then large and 
thus the Bondi-Hoyle radius,
\begin{equation}
\Rbh = 2G\ms/(v^2 + c_s^2),
\end{equation}
 becomes smaller than the tidal radius and  determines
the accretion rates.

\section{Resultant IMFs}
We can use the above formulation of the accretion rates,  with a simple model for the stellar
cluster in the two physical regimes, in order to derive the resultant mass functions (Bonnell \etal~2001b).  The primary difference is the power of the stellar mass in the accretion rate
equation, $\ms^{2/3}$ for tidal-accretion and $\ms^2$ for Bondi-Hoyle accretion. Starting from a
gas rich cluster with equal stellar masses, tidal accretion results in higher accretion rates
in the centre of the cluster where the gas density is highest. This results in a spread of stellar
mass and a mass segregated cluster. The lower dependency of the accretion rate on the stellar mass results in a fairly shallow IMF of the form (where Salpeter is $\gamma=-2.35$)
\begin{equation}
dN/d\ms \propto \ms^{-3/2}.
\end{equation}
Once the cluster core enters the stellar dominated regime where Bondi-Hoyle accretion
occurs, the higher dependency of the accretion rate on the stellar mass results in a steeper
mass spectrum. Zinnecker (1982) first showed how a Bondi-Hoyle type accretion
results in a $\gamma=-2$  IMF. In a more developed model of accretion into the core of a cluster with a pre-existing mass segregation, the resultant IMF is of the form
\begin{equation}
dN/d\ms \propto \ms^{-5/2}.
\end{equation}
This steeper IMF applies only to those stars that accrete the bulk of their mass
in the stellar dominated regime, ie. the high-mass stars in the core of the cluster (see Figure~1).

\begin{figure}[ht]
\sidebyside
{\centerline{\psfig{file=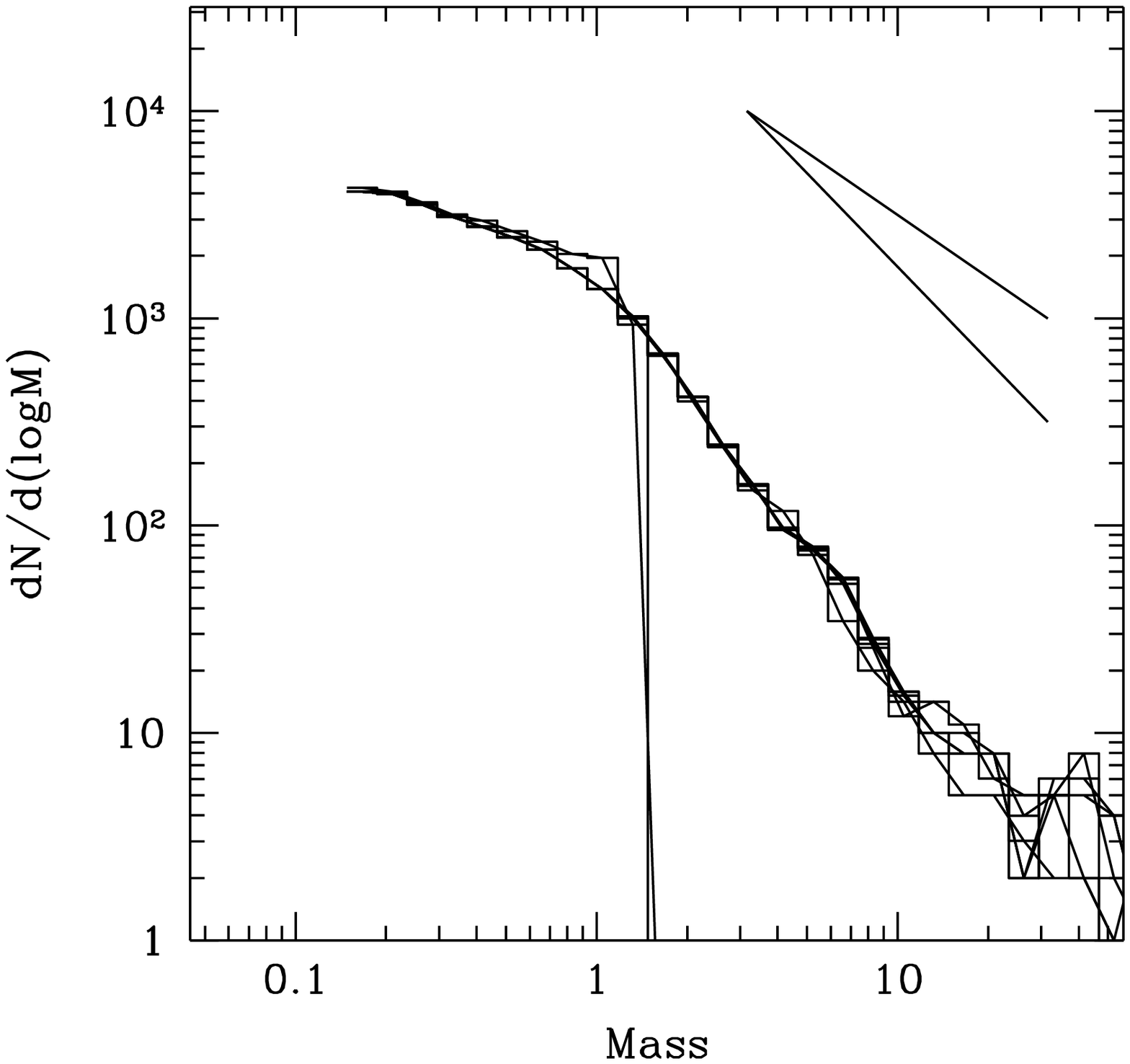,height=1.95in}}
\caption{The IMF that results from competitive accretion in a model cluster. the two
power-law IMF results from a combination of tidal accretion for low-mass stars and Bondi-Hoyle accretion for high-mass stars (BCBP~2001).}}
{\centerline{\psfig{file=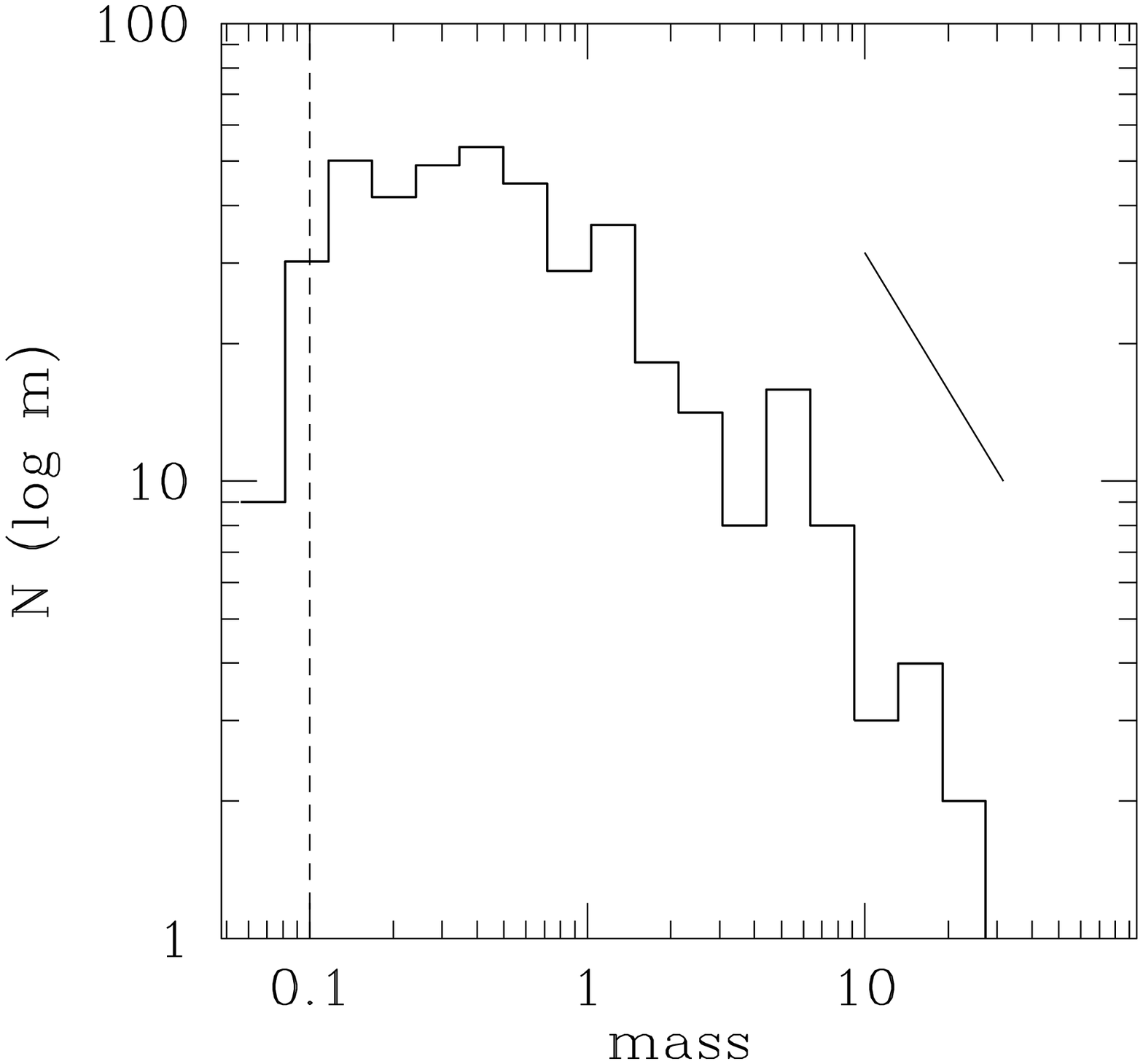,height=1.95in}}
\caption{The IMF that results from a numerical simulation of the ffragmentation of a turbuelnt cloud and the formation of a stellar cluster containing 419 stars.  
The comparison slope is for a $\gamma=-2$ IMF (BBV~2003).}}
\end{figure}

\section{The formation of stellar clusters}

In order to asses the role of accretion in determining the IMF, we need to consider
self-consistent models for the formation of the cluster. 
We followed the fragmentation of a 1000 \solmas\ cloud with a $0.5$ pc radius (Bonnell, Bate \& Vine~2003).
The cloud is initially supported by turbulence which decays on the cloud's crossing time.
As the turbulence decays, it generates filamentary structure which act as the seeds
for the subsequent fragmentation. The cluster forms in a hierarchical
manner with many subclusters forming before eventually merging into one larger cluster.
In all 419 stars form in $5 \times 10^5$ years.
The simulation produces  a field star IMF with a shallow slope for
low-mass stars steepening to a Salpeter-like slope for high-mass stars (Figure~2). The stars
all form with  low masses  and accrete up to their
final masses. The stars that are in the centres of the subclusters accrete more gas and thus
become higher-mass stars.

\begin{figure}[ht]
\centerline{\vspace{-1.75in}\psfig{file=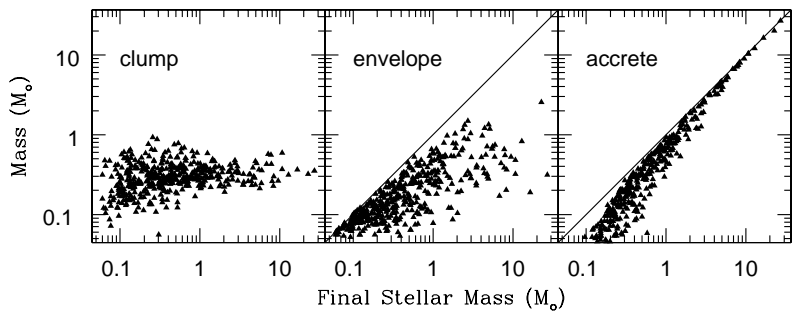,height=5in,rheight=1.9in}}
\vspace{1.75in}\caption{The relation between fragment mass (left) mass in a contiguous envelope (middle)
and mass accreted from beyond the cluster (right) is plotted against final stellar mass.
High-mass stars attain their mass due to competitive accretion of gas infalling
into the stellar cluster (Bonnell, Vine \& Bate~2004).}
\end{figure}

A careful dissection of the origin of the more massive stars reveals the importance
of competitive accretion in setting the IMF.  the Lagrangian nature of the SPH simulations
allows us to trace the mass from which a star forms. We can therefore distribute this
mass as being in one of three categories: 1) The original fragment which formed the star;
2) A contiguous envelope around the fragment until we reach the next forming star; and
3) mass which originates beyond the forming group or cluster of stars.  Figure~3
shows the three contributions to the final stellar mass of the 419 stars (Bonnell, Vine \& Bate~2004).
We see that the initial fragment mass is generally around $0.5 \solm$ and fails to account for
both high-mass and low-mass stars. The contiguous envelope does a better job as it accounts for
subfragmention into lower-mass stars. It still fails to explain the mass of higher-mass stars.
We are thus left with accretion from beyond the forming protocluster in order to explain the
existence of higher-mass stars. Thus, high-mass stars are formed due to competive accretion of gas
that {\sl infalls} into the stellar cluster. 

\section{Massive stars and cluster formation}

One of the predictions that we can extract from the simulations  is that there should
be a strong correlation between the mass of the most massive star and the number
of stars in the cluster (Figure~4, Bonnell, Vine \& Bate~2004). This occurs due to the simultaneous accretion of stars and gas into
a forming stellar cluster. Given an effective initial efficiency of fragmentation, for every star
that falls into the cluster a certain amount of gas also enters the cluster. This gas joins
the common reservoir from which the most massive star takes the largest share in this
competitive environment. Thus, the mass of the most massive star increases as
the cluster grows in numbers of stars.

\begin{figure}[ht]
\sidebyside
{\centerline{\psfig{file=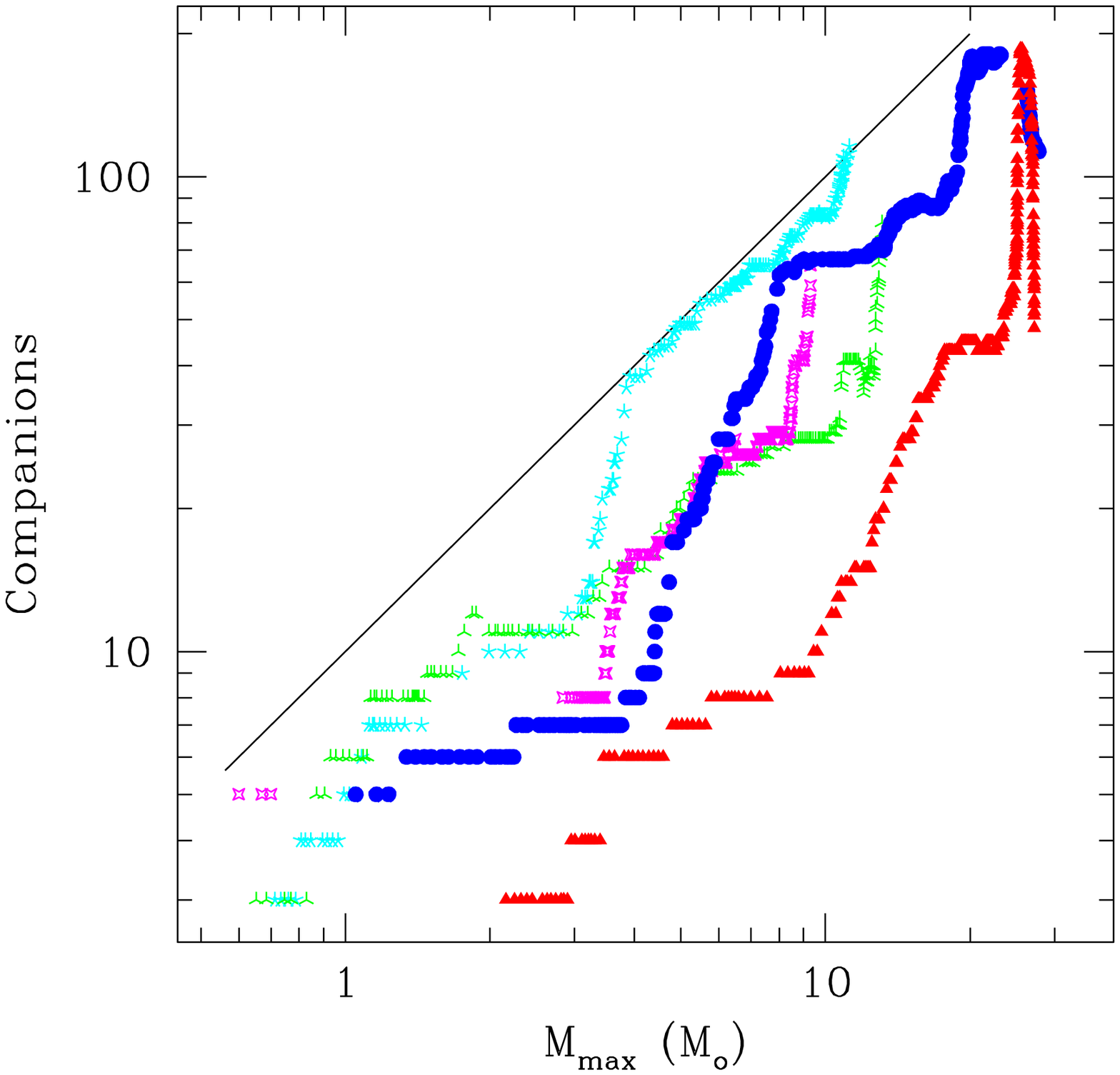,height=1.8in}}
\caption{The  number of stars in a subcluster is plotted against the mass of the
most massive star therein(BVB~2004).}}
{\centerline{\psfig{file=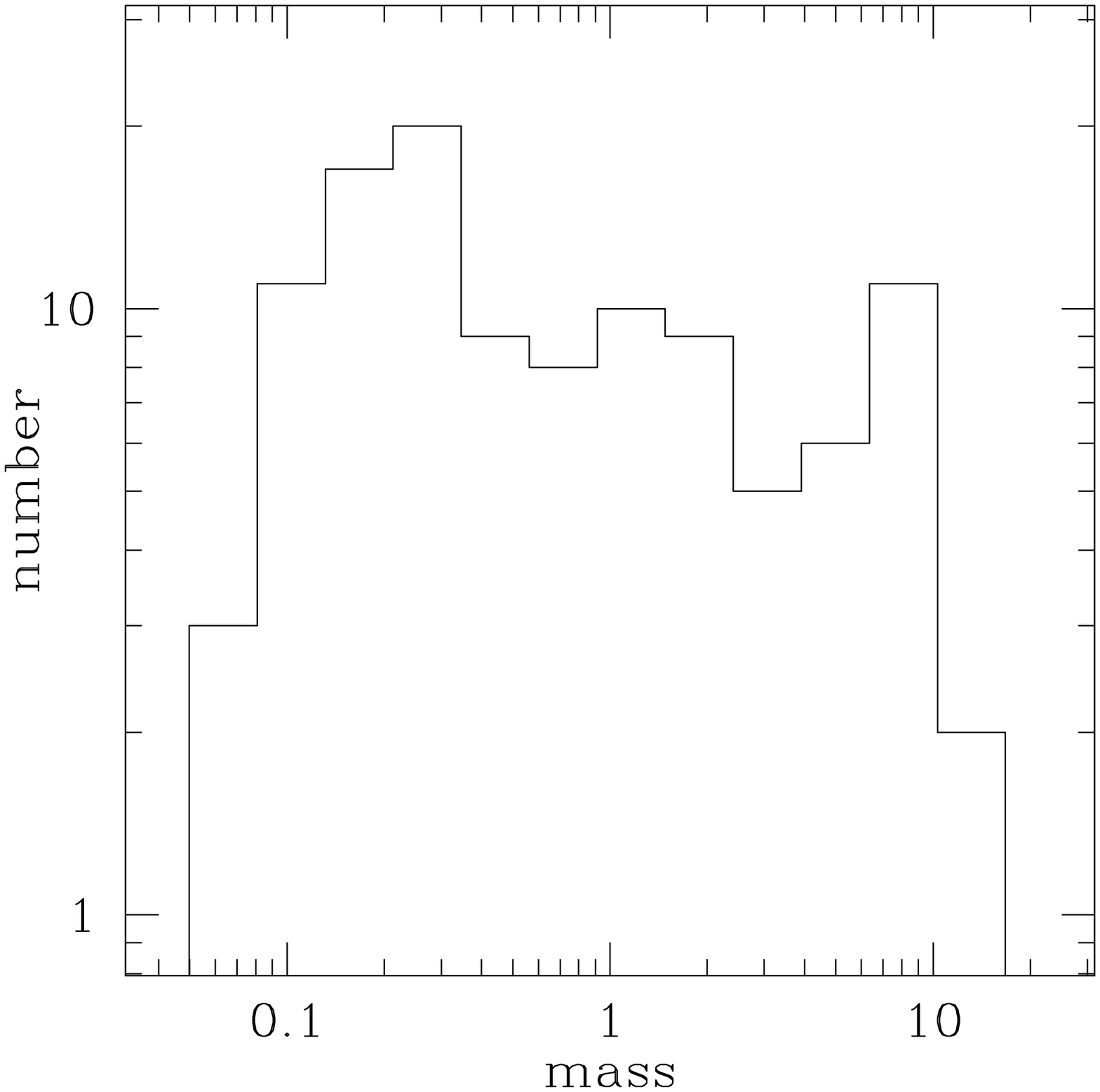,height=1.6in}}
\caption{The IMF that results from a fragmentation of a 1000 \solmas\ cloud
where the Jeans mass is $5$ \solmas..}}
\end{figure}

\section{What is the role of the Jeans mass?}

If competitive accretion is the determining factor in setting the IMF, what role does the 
Jeans mass play? We saw above that the average fragment mass in which stars
form is $\approx 0.5$ \solmas\ while the initial Jeans mass is $= 1 \solm$. Thie Jeans mass is 
therefore important at setting the scale of the initial fragmentation.
We have run a simulation identical to the one in Bonnell, Bate \& Vine~(2003) except
with a Jeans mass of $5 \solm$. The resultant IMF, shown in Figure~5, is fairly shallow up to
masses of $\approx 5-10 \solm$, corresponding roughly to the Jeans mass. In Figure~2
we see that the slope of the IMF becomes steeper at $\approx 1 \solm$, the Jeans mass in this
simulation. We can thus deduce that the Jeans mass helps set the {\sl knee} in the IMF, with lower
masses determined by fragmentation and the tidal shearing of nearby stars, while higher-masses
are determined by accretion in a clustered environment.

\section{Conclusions}

Competiitve accretion is a simple physical model that can explain the origin of the initial mass function.
It relies on gravitational competition for gas in a clustered environment and does not necessarily
involve large-scale motions of the accretors. The two physical regimes (gas and stellar dominated potentials) naturally result in a two power-law IMF.  It is probably even more important for any
model of the IMF to have secondary characteristics that can be compared to observations. For example, competive
accretion naturally results in a mass segregated cluster, and predicts that massive star formation
is intrinsically linked to the formation of a stellar cluster.  Some of the simulations reported
here were performed with the UK's Astrophysical Fluid facility, UKAFF.


\end{document}